\documentstyle[12pt]{article}

\begin{document}

\renewcommand{\theequation}{\thesection.\arabic{equation}}
\newcommand{\reseteqnum}{\setcounter{equation}{0}}

\title{
\hfill
\parbox{4cm}{\normalsize KUNS 1388\\HE(TH)~96/02\\hep-th/9605017}\\
\vspace{2cm}
On thermodynamics of black p-branes
\vspace{1.5cm}}
\author{Tomomi Muto\thanks{e-mail address:
\tt muto@gauge.scphys.kyoto-u.ac.jp}\\
{\normalsize\em Department of Physics, Kyoto University}\\
{\normalsize\em Kyoto 606-01, Japan}}
\date{\normalsize May, 1996}
\maketitle
\vspace{1cm}

\begin{abstract}
\normalsize

Thermodynamic properties of a class of
black $p$-branes in $D$-dimensions
considered by Duff and Lu
are investigated semi-classically.
For black $(d-1)$-brane,
thermodynamic quantities
depend on $D$ and $d$ only through
the combination $\tilde d \equiv D-d-2$.
The behavior of the Hawking temperature
and the lifetime vary with $\tilde d$,
with a critical value $\tilde d=2$.
For $\tilde d>2$, there remains a remnant,
in which non-zero entropy is stored.
Implications of the fact that
the Bekenstein-Hawking entropy
of the black $(d-1)$-brane depend
only on $\tilde d=D-d-2$ is discussed
from the point of view of duality.

\end{abstract}

\newpage
\section{Introduction}

It has been known that
there is a striking analogy
between the laws of black hole
and the laws of thermodynamics.
However it has raised problems such as
statistical description of the entropy
and the information loss paradox.

Recently, Strominger and Vafa calculated
the entropy of a certain extremal black hole
microscopically using D-brane technology \cite{SV}.
After that, some attempts have been made
to extend their calculation to other solutions
and non-extremal limit \cite{D-brane}.
By inspecting these investigations,
it seems that the constancy of a dilaton
is a key ingredient
for the D-brane interpretation
of the entropy to be applicable,
as pointed out in \cite{KT}.

However, various kinds of black $p$-branes
with a non-constant dilaton are known.
So, it is interesting to study
whether some microscopic explanation
of the entropy is possible
for such black $p$-branes.
As a step toward this direction,
we investigate thermodynamic properties of
a class of dilatonic black $p$-branes
which are considered by Duff and Lu \cite{DL,DKL},
semi-classically.

The action of this model includes
an antisymmetric tensor potential
of rank $d$ in $D$-dimensional spacetime,
interacting with gravity and a dilaton,
and it represents a part of the low energy
effective actions of typeIIA,
typeIIB superstrings and M-theory.
There are two types of black $p$-brane solutions
which are related by duality:
black $(d-1)$-brane with an electric charge
and black $(\tilde d -1)$-brane with a
magnetic charge.
It turns out that in the expression of
thermodynamic quantities of the $(d-1)$-brane,
$D$ and $d$ appear only through the combination
$\tilde d=D-d-2$,
hence the thermodynamic properties vary with $\tilde d$.
As the black $(d-1)$-brane approaches to the
extremal limit,
the Hawking temperature goes to zero for $\tilde d>2$,
while it goes to infinity for $\tilde d<2$.
For $\tilde d=2$, the temperature is finite
at this limit.
Thus thermodynamic property of the black
$(d-1)$-brane changes by the sign of $\tilde d-2$.

This feature also appears in lifetime of the black
$(d-1)$-brane.
By assuming that the black $(d-1)$-brane radiates
just like a black body of the Hawking temperature,
and loses its mass due to the radiation of energy,
we calculate the lifetime,
that is, the time until the mass of the
black $(d-1)$-brane decreases to the extremal limit.
It is finite for $\tilde d>2$
and infinite for $\tilde d<2$.
It can be interpreted that non-zero entropy is stored
in the remnant $(d-1)$-brane for $\tilde d<2$.

In section 2, we review the black $p$-brane
solutions discussed by Duff and Lu.
Thermodynamic properties of the black $p$-brane
are investigated in section 3.
In section 4, implications of the fact that the black
$(d-1)$-brane depend only on $\tilde d$
are discussed by taking duality into account.

\section{Black p-branes in string theory}
\reseteqnum

In this section, we recapitulate the black $p$-brane
solutions discussed in \cite{DL,DKL}.
We consider the following $D$-dimensional action
\begin{equation}
S=\frac{1}{16\pi} \int d^Dx \sqrt{-g} \left[R-\frac{1}{2}(\partial \phi)^2
-\frac{1}{2(d+1)!} e^{-\alpha \phi} F_{d+1}^2 \right]
\label{eq:action}
\end{equation}
where $\phi$ denotes a dilaton, and $F_{d+1}$ is a field strength
for an antisymmetric tensor $A_d$ of rank $d$, i.e.
\begin{equation}
F_{d+1}=d A_d.
\end{equation}
We take the constant $\alpha$ to satisfy
\begin{equation}
\alpha^2=4-\frac{2d(D-d-2)}{D-2}.
\label{eq:alpha}
\end{equation}

This choice of $\alpha$ has a following significance.
Bosonic part of low energy effective actions of
typeIIA and typeIIB superstrings are
\begin{eqnarray}
I_{\rm IIA}&=&\frac{1}{16\pi} \int d^{10}x \sqrt{-g} \left[R
-\frac{1}{2}(\partial \phi)^2
-\frac{1}{4} e^{\frac{3}{2} \phi} F_{2}^2
-\frac{1}{12} e^{-\phi} F_3^2
-\frac{1}{48} e^{\frac{1}{2} \phi} F_{4}^2 \right]+... \cr
I_{\rm IIB}&=&\frac{1}{16\pi} \int d^{10}x \sqrt{-g} \left[R
-\frac{1}{2}(\partial \phi)^2
-\frac{1}{12} e^{-\phi} F_3^{(1)2}
-\frac{1}{12} e^{\phi} F_3^{(2)2} \right]+...
\end{eqnarray}
where self-dual five-form of typeIIB theory is set to zero.
Here various kinds of field strength appear,
but the coefficient of the dilaton satisfies the equation
(\ref{eq:alpha}) for each value of $d$.
In addition, bosonic part of eleven-dimensional
supergravity, which is considered as the low energy
effective action of M-theory,
\begin{equation}
I_{\rm M}=\frac{1}{16\pi} \int d^{10}x \sqrt{-g} \left[R
-\frac{1}{12} F_{3}^2 \right]+...
\end{equation}
also contains the action (\ref{eq:action}):
in this case, $\alpha=0$, and hence $\phi$ decouples.

The equations of motion for the action (\ref{eq:action}) are
\begin{equation}
\nabla^M (e^{-\alpha \phi} F_{M M_1...M_d})=0,
\label{eq:eq1}
\end{equation}
\begin{equation}
\Box \phi +\frac{\alpha}{2(D-2)!} e^{-\alpha \phi} F_{d+1}^2=0,
\end{equation}
\begin{eqnarray}
R_{MN}=\frac{1}{2} \nabla_{M}\phi \nabla_{N}\phi
&+&\frac{2}{(D-3)!} e^{-\alpha \phi} F_{MM_1...M_d}F_N^{M_1...M_d} \cr
&-&\frac{1}{2(D-2)(D-2)!} e^{-\alpha \phi} g_{MN} F_{d+1}^2
\label{eq:eq3}
\end{eqnarray}
where $M,N=0,1,...,D-1$.

The black $(d-1)$-brane solution for these equations is
\begin{eqnarray}
ds^2&=&-\left[1-\left(\frac{r_+}{r}\right)^{\tilde d} \right]
\left[1-\left(\frac{r_-}{r}\right)^{\tilde d} \right]
^{-\frac{d}{d+\tilde d}} dt^2
+\left[1-\left(\frac{r_+}{r}\right)^{\tilde d} \right]^{-1}
\left[1-\left(\frac{r_-}{r}\right)^{\tilde d} \right]
^{-1+\frac{\alpha^2}{2 \tilde d}} dr^2 \cr
&&+r^2 \left[1-\left(\frac{r_-}{r}\right)^{\tilde d} \right]
^\frac{\alpha^2}{2 \tilde d} d\Omega_{\tilde d+1}^2
+\left[1-\left(\frac{r_-}{r}\right)^{\tilde d} \right]
^{\frac{\tilde d}{d+\tilde d}} \delta_{ij} dx^i dx^j,
\label{eq:d-1}
\end{eqnarray}
\begin{equation}
A_{01...d-1}=\left(\frac{r_+ r_-}{r^2}\right)^\frac{\tilde d}{2},
\end{equation}
\begin{equation}
e^{-2\phi}
=\left[1-\left(\frac{r_-}{r}\right)^{\tilde d} \right]^{-\alpha}
\end{equation}
where $i,j=1,2,...d-1$,
$d\Omega_{n}^2$ is the metric on unit $n$-sphere
and we define
\begin{equation}
\tilde d = D-d-2.
\end{equation}
For this solution, field strength is expressed as
\begin{equation}
e^{-\alpha \phi} *F_{d+1}
=\tilde d (r_+ r_-)^\frac{\tilde d}{2}\epsilon_{\tilde d +1}
\end{equation}
where $\epsilon_{\tilde d +1}$ is a volume form
on unit $(\tilde d +1)$-sphere.

This solution has an event horizon at $r=r_+$,
and an inner horizon at $r=r_-$ for $r_+ > r_-$.
$r_+ = r_-$ corresponds to the extremal BPS state.

This black $(d-1)$-brane has an "electric" charge
with respect to the rank $(d+1)$ field strength,
\begin{equation}
Q=\frac{1}{16\pi} \int_{S^{\tilde d+1}} e^{-\alpha \phi} *F_{d+1}
=\frac{1}{16\pi} \Omega_{\tilde d+1} \tilde d
(r_+ r_-)^\frac{\tilde d}{2}
\label{eq:charge}
\end{equation}
where $\Omega_n$ is the volume of unit $n$-sphere.

In order to obtain black $p$-branes with a
"magnetic" charge,
we use the fact that the equations of motion
(\ref{eq:eq1})-(\ref{eq:eq3}) are invariant
under the following duality transformation
\begin{equation}
d \rightarrow \tilde d,
\end{equation}
\begin{equation}
\alpha \rightarrow -\alpha,
\end{equation}
\begin{equation}
e^{-\alpha \phi} *F_{d+1} \rightarrow F_{d+1}.
\end{equation}

By applying this transformation
to the "electrically" charged black $(d-1)$-brane,
we get a "magnetically" charged black
$(\tilde d-1)$-brane,
\begin{eqnarray}
ds^2&=&-\left[1-\left(\frac{r_+}{r}\right)^d \right]
\left[1-\left(\frac{r_-}{r}\right)^d \right]
^{-\frac{\tilde d}{d+\tilde d}} dt^2
+\left[1-\left(\frac{r_+}{r}\right)^d \right]^{-1}
\left[1-\left(\frac{r_-}{r}\right)^d \right]
^{-1+\frac{\alpha^2}{2d}} dr^2 \cr
&&+r^2 \left[1-\left(\frac{r_-}{r}\right)^d \right]
^{\frac{\alpha^2}{2d}} d\Omega_{d+1}^2
+\left[1-\left(\frac{r_-}{r}\right)^d \right]
^{\frac{d}{d+\tilde d}} \delta_{ij} dx^i dx^j,
\end{eqnarray}
\begin{equation}
F_{d+1}=d (r_+ r_-)^\frac{d}{2} \epsilon_{d +1},
\end{equation}
\begin{equation}
e^{-2\phi}
=\left[1-\left(\frac{r_-}{r}\right)^d \right]^{\alpha}
\label{eq:dual d-1}
\end{equation}
where $i,j=1,2,...,\tilde d-1$.

\section{Thermodynamics of black p-branes}
\reseteqnum

In this section, we investigate thermodynamic
properties of the black $(d-1)$-brane.
In the following, we consider quantities
per unit $(d-1)$-volume for extensive ones.
%Here "unit p-volume" means the region
%whose i-th coordinate
%lies between $x^i$ and $x^i+1$.
%Here "unit $p$-volume" means the region
%whose i-th coordinate $x^i$
%takes from some value $x_0^i$ to $x_0^i+1$.
If we rewrite the metric of the "electric"
$(d-1)$-brane (\ref{eq:d-1}) as
\begin{equation}
ds^2=-A(r)dt^2 +B(r)dr^2 +r^2 C(r)d\Omega_{\tilde d+1}^2
+D(r)\delta_{ij}dx^i dx^j,
\end{equation}
ADM mass of the black $(d-1)$-brane is
calculated following \cite{Lu},
\begin{eqnarray}
M&=&-\frac{1}{16\pi} \Omega_{\tilde d+1}
\left[(\tilde d+1) r^{\tilde d+1} \partial_r C
+(d-1) r^{\tilde d+1} \partial_r D
-(\tilde d+1) r^{\tilde d} (B-C) \right]_{r \rightarrow \infty}\cr
&=&\frac{1}{16\pi} \Omega_{\tilde d+1}
\left[(\tilde d+1) r_+^{\tilde d} -r_-^{\tilde d} \right].
\end{eqnarray}
At the extremal limit $r_+=r_-$,
the mass coincides with the charge $Q$ (\ref{eq:charge}).

The Bekenstein-Hawking entropy
of the black $(d-1)$-brane is one forth of
the area $A_{\rm EH}$ of the event horizon,
\begin{equation}
S=\frac{1}{4}A_{\rm EH}
=\frac{1}{4}\left[(r^2 C)^\frac{\tilde d+1}{2}
D^\frac{d-1}{2} \right]_{r=r_+}
=\frac{1}{4}\Omega_{\tilde d+1}r_+^{\tilde d+1}
\left[1-\left(\frac{r_-}{r_+}\right)^{\tilde d} \right]
^\frac{\tilde d+2}{2 \tilde d}.
\end{equation}
(We have set the Newton constant one.)
We note that the entropy vanishes at the extremal
limit.

The Hawking temperature can be computed
by analytically continuing in $t$
and requiring that the resulting Riemannian
space has no conical singularity.
This requires a periodic identification
in imaginary time and the temperature
is inverse of the period.
For the black $(d-1)$-brane,
\begin{equation}
T=\frac{1}{4\pi} \left[\sqrt{\partial_r A
\partial_r \left( \frac{1}{B} \right)}\right]_{r=r_+}
=\frac{\tilde d}{4\pi r_+}
\left[1-\left(\frac{r_-}{r_+}\right)^{\tilde d} \right]
^{\frac{\tilde d-2}{2 \tilde d}}.
\end{equation}

Chemical potential associated with the electric
charge (\ref{eq:charge}) is the value of the gauge
potential at the horizon,
\begin{equation}
\mu=A_{01...d-1}|_{r=r_+}
=\left(\frac{r_-}{r_+}\right)^{\frac{\tilde d}{2}}.
\end{equation}

We can verify that
these quantities satisfy the first law of
thermodynamics
\begin{equation}
dM=TdS+\mu dQ.
\end{equation}

In order to obtain thermodynamic quantities
for the "magnetic" $(\tilde d -1)$-brane,
we only need to exchange $d$ and $\tilde d$.

Now we discuss thermodynamic properties
of the "electric" $(d-1)$-brane.
What is the most remarkable thing is that
in the expression of thermodynamic quantities,
"$d$" does not appear explicitly
when we write $D=d+\tilde d-2$.
We discuss this point in section 4.
Next, we can see that the behavior of the temperature
varies with the value $\tilde d$
with a critical value $\tilde d =2$.
At the extremal limit, $M=Q$,
the temperature is zero and finite
for $\tilde d >2$ and $\tilde d =2$, respectively.
For $\tilde d <2$, the temperature goes to infinity
as the mass approaches to the extremal limit.

The fact that $\tilde d =2$ is critical
also appears in lifetime of the black $(d-1)$-brane.
We assume that the black $(d-1)$-brane radiates
just like a black body of temperature $T$,
and the back reaction effect is to cause
the black $(d-1)$-brane to lose mass
at the same rate at which energy is radiated
to infinity.
To estimate the order of the magnitude
of the evaporation rate,
we consider the thermal radiation of massless fields.
The energy density of massless fields
in thermal equilibrium at temperature $T$
is proportional to $T^D$ in $D$-dimensional spacetime
(Stefan Boltzmann's law).
So the energy flux from the black $(d-1)$-brane
is proportional to $T^D A_{\rm EH}$.
If we let the deviation of the mass from
the extremal limit as $\delta M$,
the Hawking temperature $T$ and the area $A_{\rm EH}$
of the event horizon behaves as
\begin{equation}
T \sim (\delta M)^\frac{\tilde d-2}{2 \tilde d},
\end{equation}
\begin{equation}
A_{\rm EH} \sim
(\delta M)^\frac{\tilde d+2}{2 \tilde d}
\end{equation}
to leading order in $\delta M$.
Therefore $\delta M$ satisfies the following
differential equation
\begin{equation}
\frac{d(\delta M)}{dt}
\sim -(\delta M)^{\frac{\tilde d-2}{2 \tilde d}D
+\frac{\tilde d+2}{2 \tilde d}}.
\end{equation}

If $\tilde d=2$, the solution takes the form
\begin{equation}
\delta M \sim e^{-at}
\end{equation}
with some positive constant $a$, that is,
the mass decreases exponentially to the
extremal limit.
If $\tilde d \neq 2$, the solution takes the form,
\begin{equation}
\delta M \sim \left[
\frac{(D-1)(\tilde d-2)}{2\tilde d} t +b \right]
^{-\frac{2\tilde d}{(D-1)(\tilde d-2)}}
\end{equation}
with some positive constant $b$.
As $D-1$ is positive,
the behavior of $\delta M$ changes by the sign of
$\tilde d-2$.
For $\tilde d<2$, $\delta M$ becomes zero
in finite time like Schwarzschild black hole
in four dimensional spacetime.
For $\tilde d>2$, it takes infinite time
until the mass of the black $(d-1)$-brane
decreases to the extremal limit.
It can be interpreted that
there remains a remnant, in which non-zero
entropy is stored.

\section{Discussion}
\reseteqnum

In this section, we discuss implications of the fact
that thermodynamic quantities,
especially, the Bekenstein-Hawking entropy,
of the black $(d-1)$-brane depend on $d$ and $D$
only through the combination $\tilde d=D-d-2$.
In the process of the calculation, they are
rather complicated expressions in $d$ and $\tilde d$.
After the calculation, however, they depend only on
$\tilde d$.
We would like to stress that this is not so trivial.
If the spacetime is a direct product of
a $\tilde d+3$-dimensional black hole and $R^{d-1}$,
it is a natural result that quantities per unit
$(d-1)$-volume are independent of $d$.
But this is not the case.
Above all, this is a characteristic feature of the choice of
$\alpha$ (\ref{eq:alpha}).
It is reasonable to think that this fact is not
accidental but has some physical explanation.

Of course, quantum effects correct the solution
in general, since the black $(d-1)$-brane
does not corresponds to BPS states.
So we have no right to trust this solution
away from the extremal limit,
but it is difficult to resist speculating
about what it might mean if the solution
had a similar property after including quantum
corrections.

Now we take the duality into account.
As explained in section 2, there is a duality
which interchanges $d$ and $\tilde d$,
but $d$ and $\tilde d$ are not equivalent
in the sense that $\tilde d$ is a function of
$d$ and $D$, that is,
\begin{equation}
\tilde d = \tilde d (d,D).
\end{equation}

However, from the spirit of duality,
it is natural to consider that $d$ and $\tilde d$
are totally equivalent, hence the dimension of
spacetime $D$ is a function of $d$ and $\tilde d$,
\begin{equation}
D=D(d,\tilde d),
\end{equation}
where $d$ and $\tilde d$ are independent.

In other word, it is natural to consider that
$(d-1)$-brane and $(\tilde d-1)$-brane exist
at first, and the (uncompactified)
dimension of spacetime
is determined as the dimension in which
$(d-1)$-brane and $(\tilde d-1)$-brane are dual,
that is,
they are related by the Dirac quantization condition.

This way of thinking leads to a rather
strange conclusion that
the origin of the entropy of the black $(d-1)$-brane,
which depends only on $\tilde d$,
consists of its dual partner, $(\tilde d-1)$-brane.

As we noted, however, our discussion is based on
semi-classical treatment of the non-extremal
black $p$-branes.
In order to clarify that such an interpretation
have some suggestion to the understanding of
black $p$-branes,
it is necessary to take quantum corrections into
account \footnote{While this paper was prepared,
we received preprint \cite{LMPR}
which attempts to take one loop corrections into
account for dilatonic black $p$-branes.}.

Finally, we note that we can not have an
extremal black $p$-brane with non-zero entropy
even if we boost the solution to one of the
directions of the $p$-brane as in \cite{D-brane}.

\vskip 1cm
\centerline{\large\bf Acknowledgements}

I would like to thank S.~Yahikozawa
for careful reading of this manuscript.
This work is in part supported by
Soryushi Shogakukai.

\newcommand{\J}[4]{{\sl #1} {\bf #2}(19#3) #4}
\newcommand{\NP}{Nucl.~Phys.}
\newcommand{\PL}{Phys.~Lett.}
\newcommand{\PR}{Phys.~Rev.}
\newcommand{\PRL}{Phys.~Rev.~Lett.}
\newcommand{\MPL}{Mod.~Phys.~Lett.}
\newcommand{\PTP}{Prog.~Theor.~Phys.}
\newcommand{\CMP}{Comm.~Math.~Phys.}
\newcommand{\PRep}{Phys.~Rep.}

\end{document}